\documentclass{ws-rv9x6}
\usepackage{subfigure}   
\usepackage{ws-rv-thm}   
\usepackage[square]{ws-rv-van}   
\newcommand{\pythia}{{\sc Pythia}}
\newcommand{\herwig}{{\sc Herwig}}
\newcommand{\sherpa}{{\sc Sherpa}}
\newcommand{\cascade}{{\sc Cascade}}
\newcommand{\DGLAPshower}{{\it DGLAP shower}}
\newcommand{\CCFMshower}{{\it CCFM shower}}
\newcommand{\ccfm}{Ciafaloni:1987ur,Catani:1989yc,Catani:1989sg,Marchesini:1994wr}
\newcommand{\bfkl}{Kuraev:1976ge,Kuraev:1977fs,Balitsky:1978ic}
\newcommand{\dglap}{Gribov:1972ri,Lipatov:1974qm,Altarelli:1977zs,Dokshitzer:1977sg}

\makeindex

\begin{document}

 {\large Multiparton interactions, small-$x$  processes and diffraction}

\vskip 1.2 cm 

\author[F. Hautmann ]{F. Hautmann}

\address{
Rutherford Appleton Laboratory and University of Oxford\\
Elementaire Deeltjes Fysica, Universiteit Antwerpen}
\author[F. Hautmann and H. Jung]{H. Jung}
\address{Deutsches 
Elektronen Synchrotron,   Hamburg}

\vskip 1.3 cm 

\begin{abstract}
The connection between multiparton interaction, diffractive processes and saturation effects is discussed. The relation of the rise of the gluon density at small longitudinal 
momentum fractions $x$ with 
the occurrence of saturation, diffraction and multiparton interaction is being studied both experimentally and theoretically. We   illustrate  key ideas underlying recent progress, and stress the role of different 
theoretical approaches to small-$x$ QCD evolution in  investigations of 
 multiparton interactions.
\end{abstract}
\body


\vskip 1.6 cm 

\section{Introduction}\label{Intro}

The contribution of  multiple parton 
interactions (MPI)  to high-energy hadronic collisions has been considered since  the early 
days of the QCD parton model~\cite{Landshoff:1978fq,Paver:1982yp,Mekhfi:1983az,Humpert:1984ay,Paver:1984ux,Ametller:1985tp,Halzen:1986ue}.  
In the absence of a first-principle   
systematic approach to go beyond  single parton interaction in the 
framework of QCD factorization formulas, progress 
on MPI has since been driven mainly by Monte Carlo 
modeling~\cite{Sjostrand:1987su,Sjostrand:1986ep,Butterworth:1996zw,Sjostrand:2004pf} -- see 
the recent comprehensive  overview~\cite{Sjostrand:2017cdm} of MPI 
developments from the standpoint of the {\sc Pythia}  Monte Carlo event  generator. Within 
this context, experimental signals of MPI have emerged from comparison of Monte Carlo simulations with collider measurements at the 
S$p {\bar p}$S, HERA, Tevatron, LHC for 
 production of 
 multi-jets, multi-leptons, photons, heavy flavors~\cite{Knutsson:2009zz,Aaboud:2016dea,Khachatryan:2016rjt,Abazov:2015nnn,Abazov:2014fha,Aad:2013bjm,Chatrchyan:2013qza,Abazov:2011rd}. 

The relevance  of MPI for LHC 
phenomenology~\cite{Proceedings:2016tff,Astalos:2015ivw,Abramowicz:2013iva,Bartalini:2011jp}   
has spurred efforts in the last few years to investigate the theoretical basis of multiple parton scattering from the point of view of perturbative QCD and factorization~\cite{Blok:2010ge,Blok:2011bu,Blok:2012mw,Blok:2013bpa,Buffing:2017mqm,Diehl:2017kgu,Diehl:2015bca,Diehl:2011yj,Diehl:2011tt,Gaunt:2011xd,Gaunt:2009re,Manohar:2012pe,Manohar:2012jr}. Besides, methods have 
been suggested for  
 estimating the ratio of MPI 
 to single parton scattering contributions from 
data~\cite{Seymour:2013qka,Bahr:2013gkj,dEnterria:2017yhd,dEnterria:2016ids,dEnterria:2012jam,Baranov:2015nma,Baranov:2012re,Baranov:2011ch,Maciula:2017meb,Maina:2010vh,Berger:2011ep,Gaunt:2010pi,Kulesza:1999zh,Ceccopieri:2017oqe,Golec-Biernat:2015aza,Golec-Biernat:2014nsa,Treleani:2012zi,Calucci:2010wg,Blok:2015rka,Blok:2015afa,Blok:2016ulc,Blok:2016lmd}. 

The region of small longitudinal momentum fractions $x$ plays a particularly important role in the context of 
MPI~\cite{Hautmann:2016jym} because with 
decreasing $x$ parton densities 
grow, and with high parton densities  the probability for significant contributions beyond single parton  interaction increases. 
MPI may be expected to affect the detailed structure of the exclusive 
components of final states 
even when inclusive cross sections are not     influenced. Indeed, MPI signals are sought for experimentally in multi-differential cross sections 
and final-state correlations. 
Therefore, the exclusive 
structure of multi-particle final 
states associated with small-$x$ processes is particularly crucial to the discussion of MPI. 

\begin{figure}
\centerline{\includegraphics[width=11cm]{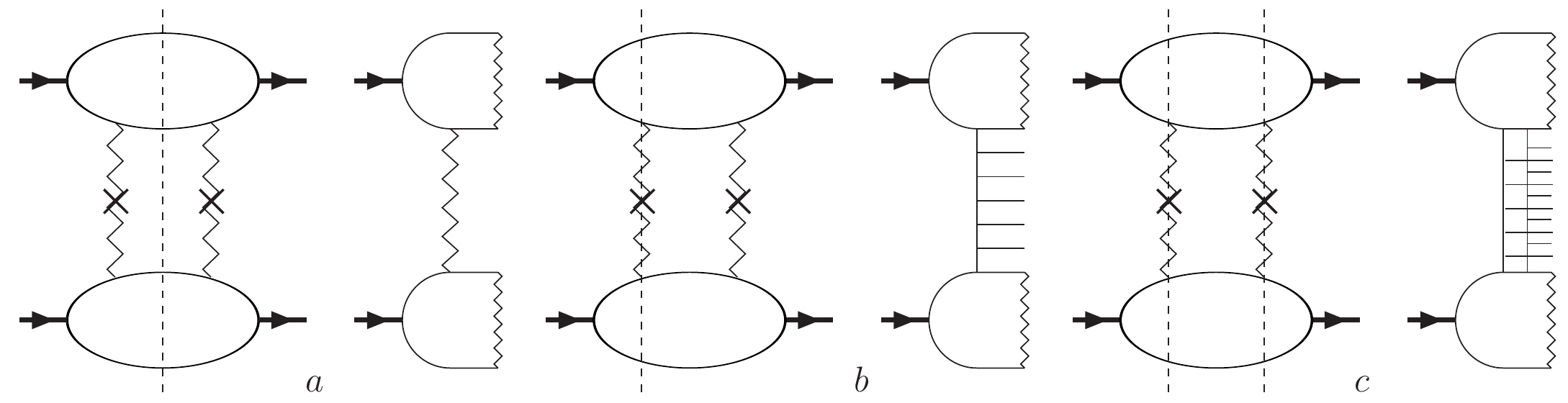}}
\caption{Sketch of Multi Parton Interactions, indicating the different ways to cut the two-pomeron diagrams: diffraction ($a$), single multiplicity (saturation) ($b$) and double multiplicity (multi-parton interaction, $c$). Figure from Ref.~\cite{Bartels:2005wa}} 
\label{smallx:agk}
\end{figure}

MPI are also naturally connected 
with aspects of small-$x$ physics 
such as diffraction and saturation. 
It is instructive to think of  this  from the viewpoint of the AGK cutting rules~\cite{Abramovsky:1973fm,Bartels:2005wa,Salvadore:2007th,Abramovsky:2011qm}  in the Regge picture of hadronic 
interactions. 
Fig.~\ref{smallx:agk} provides a schematic illustration of the relationship between diffraction, saturation and multiparton interaction 
in terms of  
Regge theory cut diagrams. The 
graph in 
fig.~\ref{smallx:agk}$a$ depicts a  
diffractive cut, while the graphs in 
fig.~\ref{smallx:agk}$b$ and 
\ref{smallx:agk}$c$,  corresponding 
to  different cuts, depict  
respectively 
 single-multiplicity interactions with 
saturation corrections and 
 double-multiplicity interactions.  
The AGK rules~\cite{Abramovsky:1973fm} connect  the different processes in 
Fig.~\ref{smallx:agk}~\cite{Bartels:2005wa}.  This connection also has  an analogue 
 in the  partonic  Monte Carlo 
models for  MPI~\cite{Sjostrand:1987su,Sjostrand:1986ep}.

In this article we give 
 a concise account of the role of  small-$x$ processes in the physics of multiparton interactions. 
We concentrate on 
general  concepts rather than 
describing specific results, with 
a view to   pointing  the 
reader to broad areas of interplay 
of small-$x$ dynamics and MPI.   
 Sec.~2 
   discusses  theoretical approaches 
to the  evolution of 
small-$x$ final states and 
their potential impact on MPI.  
 Secs.~3, 4 and  5  address  
different 
 aspects of small-$x$ physics relevant 
to MPI: diffraction, saturation and   
multi-jet production. We summarize 
 in Sec.~6.

\section{MPI and the evolution of 
small-$x$ final states}\label{sec1}

In general, experimental measurements do not allow one 
to observe multiple interactions explicitly. What  can be observed  in the 
experiment are particles and jets, distributed in phase space, and what   can be 
measured are   correlations among them as well as 
 multiplicities. Such correlations and  multiplicities 
of particles and jets can then be interpreted within 
different theoretical frameworks. 

MPI  become 
increasingly important with  energy 
as parton densities grow with 
decreasing  $x$. 
Even though they may  not 
influence    the inclusive rates 
for hard processes with a 
large $p_T$ momentum 
transfer, MPI  can contribute 
significantly to highly differential 
cross sections,  sensitive to the 
detailed distribution of multi-particle 
final states produced by parton evolution.

The evolution of parton cascades 
based  on  collinear factorization and DGLAP evolution~\cite{\dglap}, 
implemented 
in Monte Carlo event generators 
such as  
\pythia~\cite{Sjostrand:2014zea}, \herwig~\cite{Bahr:2008uq}, \sherpa~\cite{Gleisberg:2008ta} (called \DGLAPshower\ \footnote{This  terminology is used for brevity. It is a 
 misnomer though, as DGLAP is an  inclusive equation. See also note  
at the end of this section, and  
\cite{Hautmann:2017xtx} for a 
related discussion.}  in the following),  
is known 
to describe measurements well over a large range of observables.
However, 
when longitudinal momentum fractions $x$ become small and parton densities increase new effects are expected.

In fact,   new QCD dynamics   is 
 known to arise when trying to push the 
parton evolution  picture to higher and higher energies  $\sqrt{s}$. On one hand, 
soft-gluon emission 
currents~\cite{Bassetto:1984ik,Dokshitzer:1987nm} 
are modified 
by terms that depend on the total transverse momentum transmitted down the initial-state parton decay chain~\cite{Ciafaloni:1987ur}. 
Correspondingly,   
high-energy factorization 
formulas apply   which are 
 valid at fixed transverse 
momentum~\cite{Catani:1990xk,Catani:1990eg,Collins:1991ty,Levin:1991ry}. 
On the other hand, the structure of 
virtual corrections at high energy  implies, besides 
Sudakov form factors,  
transverse-momentum 
dependent (TMD) 
 -- but universal --  splitting  functions and new ``non-Sudakov" 
form factors~\cite{Ciafaloni:1987ur,Catani:1989sg,Catani:1994sq},  
 which are necessary to 
take into account soft-gluon 
coherence not only for 
collinear-ordered emissions but also in the non-ordered region that opens up at high  $\sqrt{s} / p_T$. 
These finite TMD  
corrections to parton branching 
are implemented in the 
CCFM evolution equations\cite{\ccfm}, 
and,  in the high-energy factorization  framework,  are found to have important implications for 
multiplicity distributions and the  structure of angular correlations  in 
 final states with high multiplicity~\cite{Marchesini:1992jw,Hautmann:2008vd,Hautmann:2008rd,Blok:2017wzo}.
 The CCFM evolution equations may be 
thought of as forming a bridge 
between the 
small-angle DGLAP\cite{\dglap} and 
high-energy BFKL\cite{\bfkl} regimes. 

In phenomenological analyses 
  which 
perform comparisons of   
experimental measurements  
for multi-particle final states and 
correlations with Monte Carlo 
calculations  based on \DGLAPshower\ 
event generators such as 
 \pythia, \herwig, \sherpa,  
  it is found that 
 MPI   are needed to describe  measurements  such as  soft particle spectra in minimum bias events, the underlying events in jet production as well as (de-)correlations in multi-jet 
events, including  4-jet, $b\bar{b}+j j$, $W+j$ events. 

However, in calculations based on 
 high-energy factorization~\cite{Catani:1990xk,Catani:1990eg}
and CCFM\cite{\ccfm}  
(or BFKL\cite{\bfkl})  transverse momentum dependent parton densities (unintegrated parton density 
functions uPDFs),   
the correlation between partons and particles in the final state is different from those predicted by the \DGLAPshower ,
since finite transverse momenta in the initial state are included from the beginning. The small-$x$ behavior of the (unintegrated) parton densities obtained from  CCFM evolution (e.g. Ref.~\cite{Hautmann:2013tba})  is different compared to the one from DGLAP distributions, and therefore the amount of multiple partonic interactions might also be different.
Multi-parton radiation in CCFM evolution is allowed in an 
angular-ordered region of phase space,  which is  determined by small-$x$ gluon coherence~\cite{Hautmann:2008vd}.   This 
makes a new  scenario possible,  in which 
the higher transverse momenta in the parton cascade, compared to a 
Monte Carlo simulation based on \DGLAPshower , 
can  produce final states 
 similar to what is obtained from  MPI.

The Monte Carlo event generator  \cascade \cite{Jung:2001hx,Jung:2010si}, based on CCFM uPDFs and high-energy factorization, includes initial state parton showering according to the CCFM evolution equation (called \CCFMshower\ \footnote{Unlike 
DGLAP, CCFM is an exclusive 
equation. The  terminology of   
\DGLAPshower\  and 
\CCFMshower\  is somewhat 
misleading. Strictly speaking,  
only the latter  is  defined.} 
in the following) as well as final state parton shower and hadronization. 
Studies~\cite{Deak:2011ga,Deak:2011gj,Hautmann:2012xa} of jet production 
at high  rapidity 
based on simulations with \cascade\ (without MPI) show that the energy flow outside the jets  is significantly larger than what is obtained from  \pythia\  without MPI.
In Ref. \cite{Kotko:2016lej} the relation between Monte Carlo 
simulations using MPI and high-energy factorization has been studied using mini-jets. 
It was found that jet distributions are similar in both approaches, suggesting that 
at least part of the effects that are attributed to MPI when comparing Monte Carlo calculations based on \DGLAPshower\ 
with data are already contained in the single scattering when using uPDFs and high-energy factorization.

\section{Diffractive dissociation  and MPI}\label{Diffraction}

In the scattering matrix formalism~\cite{Good:1960ba}  
hadronic diffraction is thought of as 
arising  from fluctuations in the scattering amplitude. 
Regge theory incorporates 
high-mass diffraction  through  pomeron exchange 
with triple (and multiple) pomeron 
couplings~\cite{Mueller:1970fa,Detar:1971gn,Kaidalov:1973et}. 
Both these ideas were given 
partonic interpretations, respectively in \cite{Miettinen:1978jb}
and \cite{Ingelman:1984ns}. The partonic interpretation 
of hard diffraction  leads to the 
notion of diffractive parton densities.   

Diffractive processes in deep inelastic scattering (DIS) have been  measured 
in great detail at HERA.  Based on the factorization theorem for diffractive 
DIS~\cite{Grazzini:1997ih,Berera:1995fj,Hautmann:1999ui,Collins:1997sr,Collins:2001ga}, 
diffractive parton densities~\cite{Aktas:2006hy} have been  obtained. However, using 
these diffractive parton densities   diffractive jet production in hadron-hadron 
collisions is predicted to be of the order of 10 times larger than the 
measurements~\cite{Bjorken:1992er,Affolder:2000vb,Chatrchyan:2012vc,Aad:2015xis}.  
Therefore, whilst  deep inelastic diffraction 
can  be understood in terms of   color transparency 
(see~\cite{Hautmann:2000pw,Hautmann:1998xn,Bartels:1998ea,Buchmuller:1998jv,Ingelman:2015qrt} for structure functions   
and~\cite{Hautmann:2002ff,Hautmann:2001wk} for jet and charm production), 
new   (absorptive)  
effects  occur in hadron-hadron diffraction. 
In most  of the current  phenomenological models, these are embodied in 
gap suppression factors, or gap survival probabilities. 

\begin{figure}
\centerline{
\includegraphics[width=11cm]{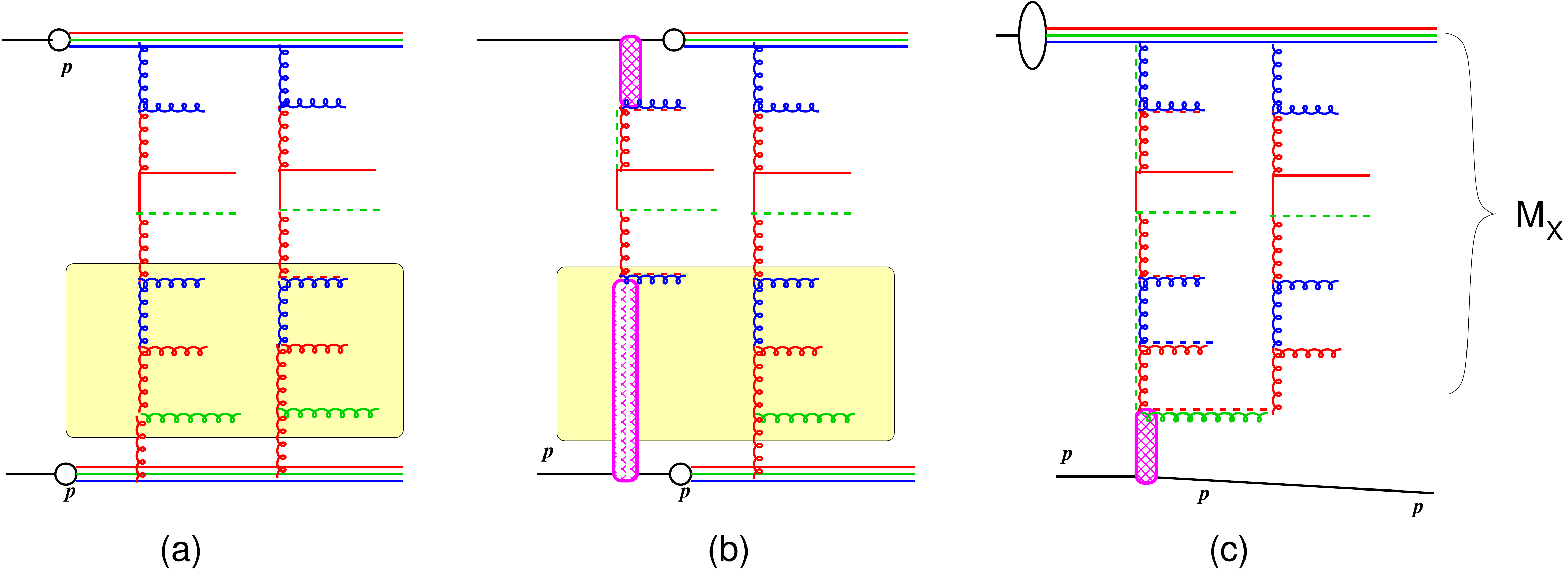}}
\caption{Left: Non-diffractive multiparton interactions.
Middle: Sketch of a diffractive process together with a non-diffractive process, 
destroying the rapidity gap. Right: multiparton interactions inside the dissociative system} 
\label{smallx:mpi}
\end{figure}

Multiparton interactions  in addition to the diffractive process can destroy 
the experimental signature of diffractive interactions, i.e.  a region in 
rapidity which is devoid of energy deposition (rapidity gap, see fig.~\ref{smallx:mpi}$b$). 
Estimates on the probability of multiple interactions lead to a suppression 
factor which is in the correct order of magnitude. Recently a relation between 
multiple parton scattering and hard diffractive processes has been implemented 
in \pythia ~\cite{Rasmussen:2015qgr}. This  model, based on diffractive parton 
densities (as obtained from HERA) together with MPI,  allows one 
to predict hard diffractive processes in hadron-hadron collisions without introducing 
artificially gap suppression factors.   
It is very encouraging that predictions coming from this new model~\cite{Rasmussen:2015qgr} 
follow the trend of the measurements of hard diffractive  dijet production at the LHC.   

In a process as shown in fig.~\ref{smallx:mpi}$a$, the density of partons in the 
forward or backward regions (indicated by the yellow box) is different from those 
of processes shown in fig.~\ref{smallx:mpi}$b$, because one interaction chain 
allows no particle radiation, while the other does. Experimentally it is difficult to 
identify such processes, but they play a role when adjusting parameters to 
describe the underlying event and multiparton interactions. It will be important 
that in forthcoming Monte Carlo tuning  diffractive processes are included, since 
this might lead  to different parameters describing particle multiplicity.

While multiparton interactions can destroy a rapidity gap, multiparton interactions 
can also occur within the diffractive dissociation system (fig.~\ref{smallx:mpi}$c$), if 
the diffractive system has a large enough mass $M_X$. At MPI@LHC 2017
the effect of MPI inside the diffractive system $M_X$ was shown for the first 
time \cite{MPIatLHC2017:grados} in a measurement of $dn/d\eta$ for event samples enhanced with single 
diffractive dissociation. 
Investigating such processes (with hard diffractive events), MPI can be studied in 
a very new environment and one can investigate in detail the energy dependence 
of MPI (and its energy dependent parameters) as a function of the center of mass    energy  
of the pomeron-proton system $M_X$ (indicated in fig.~\ref{smallx:mpi}$c$).

A further mechanism relating inelastic diffraction and MPI  is  discussed 
in Refs.~\cite{Blok:2016ulc,Blok:2016lmd,Blok:2017alw},     
 based on  two-pomeron contributions  to the  generalized double 
parton distribution  function~\cite{Blok:2012mw}.

\section{Saturation effects and MPI}\label{Saturation}

In the context of  total hadron-hadron 
cross sections~\cite{Amaldi:1979kd} and their rise with energy~\cite{Donnachie:1992ny},  
unitarity constraints  are given, at fixed impact parameter,  by the  
``black disc" limit of the scattering matrix~\cite{Heiselberg:1991is,Frankfurt:1994hf}. 
In  this limit, 
 the elastic cross section equals half the total cross section. 
Such saturation effects are 
important, for instance,  for the observed behavior of the 
diffractive cross section~\cite{Alberi:1981af}.   
In the Regge theory picture, saturation implies the  
breakdown of the simple Regge pole behavior  
and the onset of multiple pomeron exchange~\cite{Kaidalov:1986zf}. 

Unitarity arguments are also used to treat  saturation of parton densities 
in the case of hard processes 
in the limit of  high energies (or  large nuclei)~\cite{Gribov:1984tu,Mueller:1985wy,McLerran:1993ka,GolecBiernat:1998js,Mueller:1999wm},    
 with corresponding nonlinear evolution   equations~\cite{Balitsky:1995ub,Kovchegov:1999yj,JalilianMarian:1997gr,Iancu:2000hn,Gelis:2010nm}. 
This   saturation formalism    
is   given 
in terms of color-dipole degrees of freedom  rather than parton degrees of freedom. 
(See also~\cite{Flensburg:2011kj,Flensburg:2011kk} for an alternative, dipole-based perspective on saturation.)   It has    limited 
applicability in  the region of  high    transferred momenta well above the saturation scale.   
See e.g.~discussion in 
\cite{Hautmann:2007cx,Hautmann:2006xc}.  
Therefore, even though the saturation formalism  is constructed 
in the case of hard processes, it is 
generally not 
relevant  to jet physics at high $p_T$. 

On the other hand, 
for energies high enough the scale at which saturation effects become important can rise into 
 the 
region of transverse momenta $ p_T $ large compared to $\Lambda_{\rm{QCD}}$, say, 
$ p_T \sim {\cal O } ( 5 \  {\mbox{GeV}} ) $ --  potentially, a region of weak-coupling but 
nonperturbative dynamics, 
sitting  just above the saturation 
scale.  
This was the scenario proposed in  Ref.~\cite{Grebenyuk:2012qp}. 
It was emphasized  in~\cite{Grebenyuk:2012qp} that in   
this  scenario   a  connection 
arises  between saturation 
and MPI.  The connection 
could be seen in terms of  two 
aspects of the  Monte Carlo 
model~\cite{Sjostrand:1987su,Sjostrand:1986ep,Sjostrand:2004pf}: the contribution of multiple 
interactions, and the screening of 
the jet production cross section. 

One of the arguments leading to the introduction of 
MPI   
in Monte Carlo event generators~\cite{Sjostrand:1987su,Sjostrand:1986ep,Butterworth:1996zw}  
was the observation that the $2\to2$ partonic cross section integrated above $p_{T\, min}$ over the transverse momentum of the final partons can become larger than the inelastic $pp$ cross section for values of  $p_{T\,min}\gg \Lambda_{\rm{QCD}}$. Since 
the $2\to2$ cross section is a jet  --  and not an event -- 
 cross section, the ratio $\sigma(p_T > p_{T\,min})/\sigma_{inel}$ was interpreted as the average number of interactions per event. 

The $2\to2$ partonic cross section for small transverse momenta is $d\sigma/d p_T \sim \alpha_s^2/p_T^4$, and diverges for $p_{T} \to 0$.  This behavior is related to the non-perturbative nature of the process for small $p_{T}$ and is regulated in models based on collinear factorization by introducing an additional cut-off parameter, such that $d\sigma/d p_T \sim \alpha_s^2/p_T^4 \to \alpha_s^2(p_T^2+p^2_{T\,0})/(p_T^2 + p_{T\,0}^2)^2$, with an arbitrary but finite parameter $p_{T\,0}$. This parameter $p_{T\,0}$ separates the perturbative from the non-perturbative regions and has to be determined from fits to experimental data. At LHC energies, this parameter was determined to be of the order of 2-3 GeV, far above $\Lambda_{\rm{QCD}}$. In general, $p_{T\,0}$ depends on the center-of-mass 
energy $\sqrt{s}$. 

In Ref.\cite{Grebenyuk:2012qp} it was proposed to measure the leading mini-jet or leading track cross section for a direct investigation of the behavior of the cross section from the perturbative to the 
nonperturbative region at small $p_T$. In Ref. \cite{Khachatryan:2015fia} a measurement of leading track and leading mini-jet cross sections was reported which shows directly how the cross section is saturated at small $p_T$. The observation of the turnover 
of the cross section from a $\sim 1/p_T^4$ behavior to a constant value,  independent 
of Monte Carlo modeling,  is an important result. 

For interpretation of this result, 
see   discussions in  
Refs.~\cite{Rogers:2009ke,Rogers:2008ua},   
which examine impact-parameter unitarity constraints for  minijets, and in Refs.~\cite{Kutak:2008ed,Bacchetta:2010hh,Kutak:2011rb,Kutak:2011fu,Kutak:2012rf,Kotko:2015ura},  
which 
 use nonlinear evolution equations for parton distributions at fixed transverse momentum.

\section{MPI in multi-jet production}\label{Smallx}
\begin{figure}
\centerline{
\includegraphics[width=10cm]{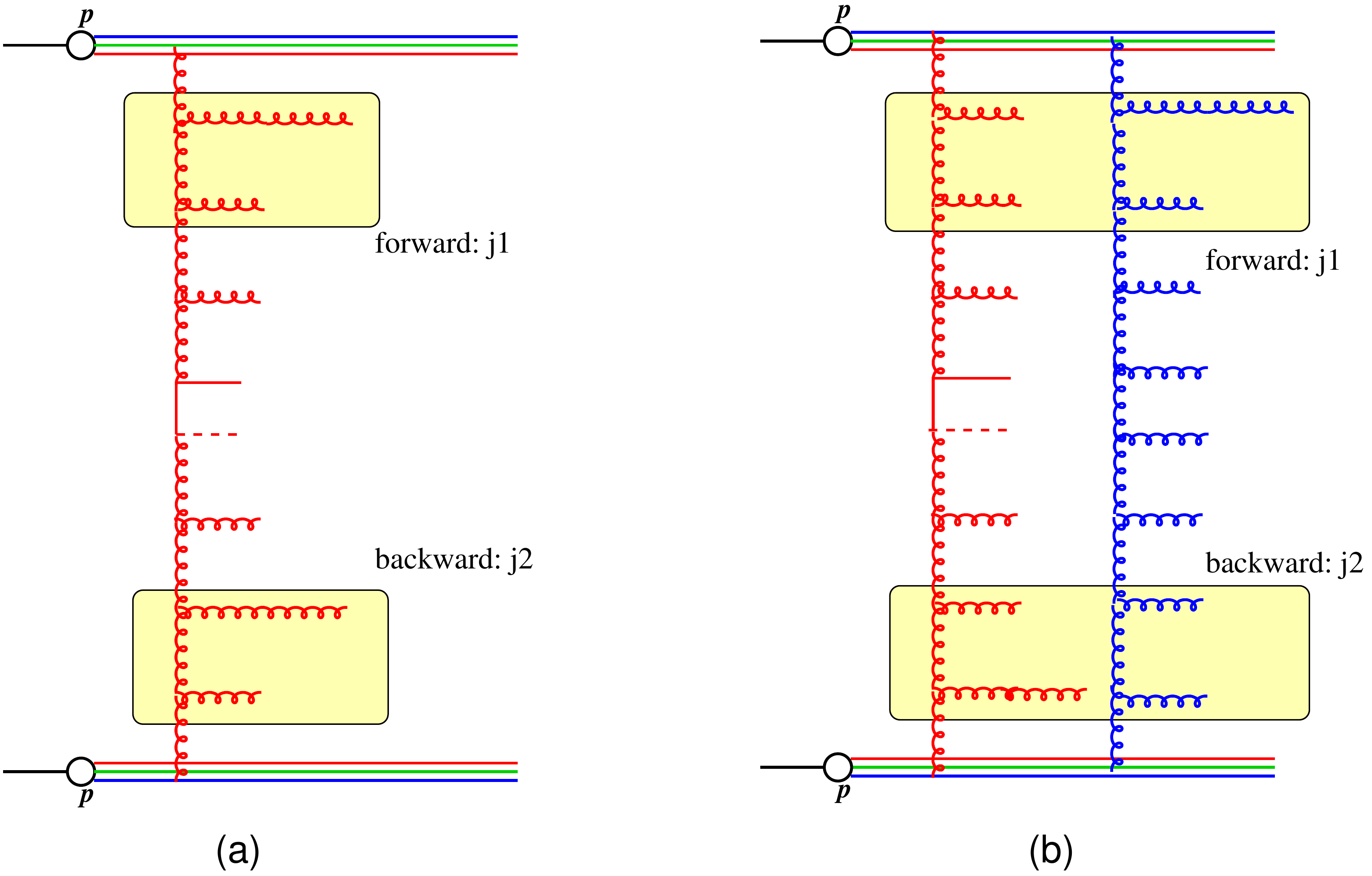}}
\caption{a: forward - backward jet production. b: forward - backward jet production from MPI} 
\label{smallx:fwd-bwd-jets}
\end{figure}

Mueller-Navelet jets~\cite{Mueller:1986ey},   having 
 comparable transverse momenta and  large rapidity separation  (fig.~\ref{smallx:fwd-bwd-jets}$a$),  
have long been investigated as a  probe of BFKL\cite{\bfkl}   dynamics. 
While  experimental measurements of forward-backward jets at the LHC  \cite{Khachatryan:2016udy,Chatrchyan:2012pb,Aad:2014pua,Aad:2011jz} 
do not point to  any striking BFKL signature,  and  \pythia\ Monte Carlo simulations  compare 
well with data,  interest has grown  in  examining   MPI contributions 
 (included in the  \pythia\ simulations) to such processes 
 (fig.~\ref{smallx:fwd-bwd-jets}$b$). Studies 
of the simplest term  from double parton scattering 
 in a collinear framework \cite{Maciula:2014pla} and  
 a transverse momentum dependent framework \cite{Ducloue:2015jba} suggest that the 
latter leads to smaller double-scattering contribution.  

A simulation, including a full treatment of the final state, based on the \CCFMshower\ will be of great importance.
The  \cascade\ Monte Carlo event generator (based on CCFM parton showers) 
is not designed at present to treat this process, since this process involves flavor channels not yet 
included in the Monte Carlo. 
Very recently a first step towards a determination of $k_T$-dependent parton densities for  all flavors has been reported in 
Refs.~\cite{Hautmann:2017xtx,Hautmann:2017fcj}, which could be used to obtain a more complete simulation of the final state in \cascade .

\begin{figure}
\centerline{
\includegraphics[width=3.5cm]{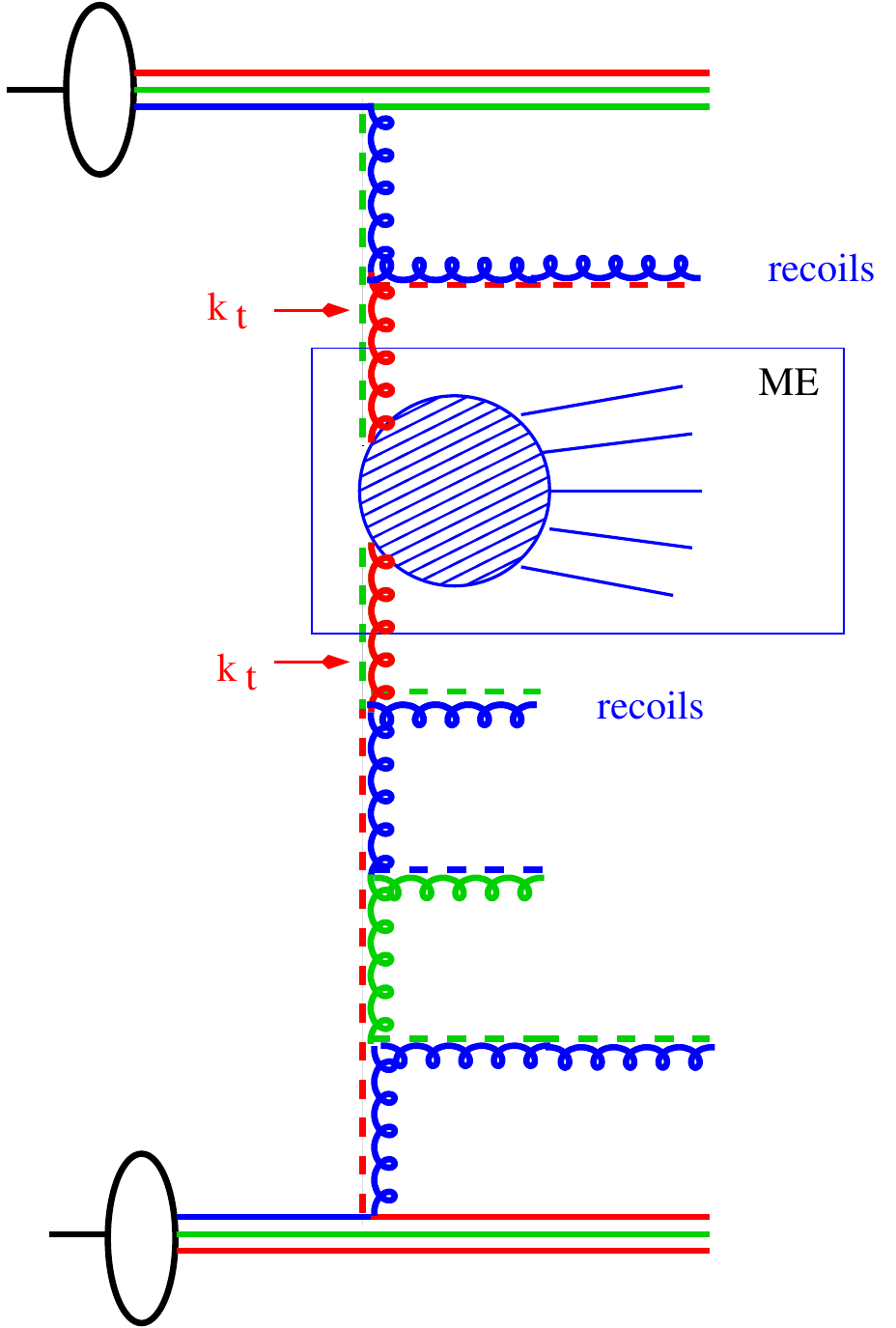}}
\caption{Sketch for recoil treatment in high-energy factorization} 
\label{smallx:recoils}
\end{figure}

At present, 
experimental  signals  for hard double parton scattering come from studies of correlations in 4-jet, $b\bar{b}$+2-jet and $W/Z$+jet events. In the case of double parton scattering, the jets or $b\bar{b}$ 
pair   or $W/Z$ bosons which come from one interaction will be decorrelated from those coming from the second interaction. However, such a decorrelation could also originate from significant transverse momenta of the initial-state partons within a single interaction. This can be studied by using multi-leg matrix element calculations (for example at least $2\to 6$ for the 4-jet case in order to allow initial transverse momenta of the $2 \to 4$ subprocess) or by using, in the 
high-energy factorization framework~\cite{Catani:1990xk,Catani:1990eg}, 
 off-shell matrix elements for $2\to4$ together with uPDFs.   One such  study has been performed 
in Ref.\cite{Kutak:2016mik} and an interesting observation was made: the prediction from a single chain interaction  for the azimuthal angular correlation  $\Delta S$  between two jet pairs  is already very close to the experimental measurements of Ref.\cite{Chatrchyan:2013qza}. 

An important point in interpreting this result, however, concerns the role of parton showers.  
In fig.~\ref{smallx:recoils} a schematic picture is given for a generic hard process described by high-energy factorization. Since the transverse momenta of the incoming partons can have any kinematically allowed value, and especially the $k_T$ can be larger than the $p_T$ of the partons of the matrix element process (indicated by ME in fig.~\ref{smallx:recoils}), the simulation of the recoils 
(in form of an explicit parton shower) 
cannot be neglected as they might significantly contribute. This is different from  the case of simulations  based on a \DGLAPshower , where the  ME-partons constitute the partons with the highest $p_T$.   Predictions  for jets based on approaches including the initial-state transverse momentum should 
  take into account   the   recoils besides the   uPDFs.  Thus, 
before a  conclusion on the size of MPI in the 
high-energy factorized calculation can be drawn, a full simulation including parton showers is   
needed.

\section{Summary}
At high energies, when the density of partons is large, multiple partonic interactions can occur with 
non-negligible rates. The relation of the rise of the parton density at small 
longitudinal momentum fractions 
$x$ and the occurrence of saturation, diffraction and multi-parton interaction is being studied both experimentally and theoretically.  

MPI affect primarily highly differential cross sections and the detailed distribution of 
multi-particle final states produced by parton evolution. 
Theoretical predictions for MPI observables are thus sensitive to the theory of final states associated with small $x$. 
We have discussed the role of different approaches to  small-$x$ QCD evolution  in the 
interpretation of measurements in terms of multi-parton interaction. 

Small-$x$ aspects of MPI influence both soft-interaction and hard-interaction processes. 
We have illustrated this by discussing their effects   in diffraction, saturation, multi-jet production.

\section*{Acknowledgments}  
We are grateful to  the editors for careful reading of the manuscript and 
advice. We thank   M.~Strikman and J.~Gaunt for useful discussions.

\bibliographystyle{ws-rv-van}
\bibliography{MPISmxDiffr5dec}


\end{document}